# Programmable self-assembly of core-shell ellipsoids at liquid interfaces

*Jack Eatson*[a,*], *Susann Bauernfeind*[b,c,*], *Benjamin Midtvedt*[d], *Antonio Ciarlo*[d], *Johannes Menath*[c], *Giuseppe Pesce*[d,e], *Andrew B. Schofield*[b], *Giovanni Volpe*[d], *Paul S. Clegg*[b], *Nicolas Vogel*[c], *D. Martin. A. Buzza*[a], *Marcel Rey*[b,d,f]

[a] Department of Physics and Astrophysics, G. W. Gray Centre for Advanced Materials, University of Hull, Hull HU6 7RX, United Kingdom;

[b] School of Physics and Astronomy, The University of Edinburgh, Peter Guthrie Tait Road, Edinburgh EH9 3FD, UK.

[c] Institute of Particle Technology (LFG), Friedrich-Alexander-Universität Erlangen-Nürnberg (FAU), Cauerstrasse 4, 91058 Erlangen, Germany

[d] Department of Physics, University of Gothenburg, SE-41296, Gothenburg, Sweden

[e] Dipartimento di Fisica "Ettore Pancini", Università degli Studi di Napoli Federico II, Naples, Italy.

[f] University of Münster, Institute of Physical Chemistry, Corrensstr. 28/30, 48149 Münster, Germany

**Abstract**

Ellipsoidal particles confined at liquid interfaces exhibit complex self-assembly behaviour due to quadrupolar capillary interactions induced by meniscus deformation. These interactions cause particles to attract each other in either tip-to-tip or side-to-side configurations. However, controlling their interfacial self-assembly is challenging because it is difficult to predict which of these two states will be preferred. In this study, we demonstrate that introducing a soft shell around hard ellipsoidal particles provides a means to control the self-assembly process, allowing us to switch the preferred configuration between these states. We study their interfacial self-assembly and find that pure ellipsoids without a shell consistently form a "chain-like" side-to-side assembly, regardless of aspect ratio. In contrast, core-shell ellipsoids transition from "flower-like" tip-to-tip to "chain-like" side-to-side arrangements as their aspect ratios increase. The critical aspect ratio for transitioning between these structures increases with shell-to-core ratios. Our experimental findings are corroborated by theoretical calculations and Monte Carlo simulations, which map out the phase diagram of thermodynamically preferred self-assembly structures for core-shell ellipsoids as a function of aspect ratio and shell-to-core ratios. This study shows how to program the self-assembly of anisotropic particles by tuning their physicochemical properties, allowing the deterministic realization of distinct structural configurations.

**Introduction**

When particles adsorb to a liquid interface, they reduce the direct contact between the two phases, lower the overall free energy of the system,[1] and provide mechanical strength to the interface due to jamming.[2,3] This can lead to the formation of stable foams,[4–7] emulsions,[8–13] bijels[14,15] or liquid marbles,[16,17] with a wide range of applications, including in the food industry,[8,9] cosmetics,[18] and medical fields.[19–22] Furthermore, the interfacial assembly can be transferred onto solid substrates to obtain nanoscale surface patterns with high fidelity over macroscopic areas.[23] These patterns are exploited in photonic[24] and phononic[25] applications, and serve as a template for the fabrication of more complex plasmonic nanostructure-[26,27] and nanowire arrays.[28,29]

Liquid interfaces further confine colloidal particles to two dimensions, enabling the study and control of their self-assembly behaviour. Monodispersed spherical colloidal particles typically self-assemble into hexagonal structures, influenced by the balance of attractive capillary and van-der-Waals forces and repulsive electrostatic and dipole forces.[30,31] For rough or nonspherical particles, strong lateral capillary interactions arise due to interface deformation caused by an undulating contact line or particle shape effects.[32–37] These dominant capillary interactions between the particles not only enhance the stability of foams[38] or emulsions,[39–41] possibly through strong effects on the interfacial rheology,[39] but also dictate their self-assembly behaviour at liquid interfaces.[34,37,42]

For ellipsoidal particles at liquid interfaces, the deformation of the meniscus along the long axis differs from that along the short axis, resulting in capillary forces with quadrupolar symmetry that greatly exceed the thermal energy $k_\mathrm{B}T$.[34] The saddle-like distortion field around the ellipses causes them to attract each other either tip-to-tip or side-to-side, while repelling each other in the side-to-tip configuration.[33,34,37]

Both tip-to-tip and side-to-side configurations have been found experimentally. At the oil/water interface, sterically-stabilized ellipsoids self-assembled into a side-to-side configuration, corroborated by calculations and simulations as the energetically most stable configuration.[43,44] Selectively removing the steric stabilizer from the tip of the ellipsoids changed the self-assembly into tip-to-tip configurations.[45] The self-assembly behaviour of charge-stabilized ellipsoids is less understood. At an air/water interface, side-to-side[46–49] and tip-to-tip[34,46,49,50] arrangements have been reported. In one study, ellipsoids with low aspect ratios assembled preferentially side-to-side, while a tip-to-tip arrangement was found for higher aspect ratios.[46] Another study reported "flower-like" tip-to-tip arrangements at the air/water interface, whereas the same particles formed a "chain-like" tip-to-tip arrangement at a decane/water interface.[50] This difference was rationalized by the different contact angle of the ellipsoids adsorbed at the air/water and decane/water interface, respectively.[50] Additionally, core-shell ellipsoids consisting of an incompressible core and a hydrogel shell initially self-assembled into a side-to-side assembly, which transitioned over time into a tip-to-tip assembly.[51] Furthermore, the shell thickness of these core-shell ellipsoids was reported to affect their assembly, with thicker shells resulting in a tip-to-tip arrangement, while thinner shells led to a side-to-side assembly.[52] To summarize, both tip-to-tip and side-to-side arrangements for seemingly similar ellipsoids have been observed experimentally. However, the origin of the preference for either structure is not yet fully understood and deterministic control of their self-assembly thus remains challenging.

A convenient and frequently used method to produce ellipsoidal particles with controlled aspect ratios is the thermo-mechanical stretching technique.[53,54] However, we recently discovered that in this fabrication process, the cleaning protocol significantly impacts the behaviour of ellipsoidal particles due to residual polymer chains that remain on their surfaces.[55] These polymer residues impact the behaviour of both the individual particle and the particle dispersion. Contrary to the previous belief that the anisotropic shape of ellipsoidal particles prevents the coffee ring effect,[56] our findings demonstrated that it is actually the presence of polymeric residues adsorbed on the particle surface that leads to homogeneous drying.[55] When these polymer residues are properly removed using an appropriate solvent,[47] the ellipsoidal particles exhibit the coffee ring effect.[55] We therefore hypothesize that the potential presence of polymer residues, an artifact of the fabrication, or cleaning process, may similarly affect the interfacial self-assembly and thus explain the variety of different assembly structures reported for ellipsoidal particles.

Here, we experimentally and computationally investigate the interfacial self-assembly behaviour of core-shell ellipsoids, focussing on the role of the shell provided by water-soluble polymer chains adsorbed to the particle surface. By carefully controlling the fabrication process and cleaning protocol, we synthesize either pure ellipsoids or core-shell ellipsoids with polymeric hairs on their surfaces. These polymers significantly influence the self-assembly behaviour of the ellipsoids. Pure ellipsoids consistently assemble into a side-to-side arrangement, independent of their aspect ratio. In contrast, core-shell ellipsoids exhibit a transition from a tip-to-tip assembly at lower aspect ratios to a side-to-side assembly at higher aspect ratios, with the onset of this transition shifting to higher aspect ratios as the relative shell size (i.e., the ratio of shell to core) increases. We support our experimental observations with theoretical calculations and Monte Carlo simulations, mapping out the phase diagram for core-shell ellipsoids as a function of aspect ratio and shell thickness, which accurately reproduces the experimental results.

## Results and Discussion

**The interfacial self-assembly behaviour of ellipsoids is influenced by their fabrication and cleaning procedures.** We compare the self-assembly of polymeric polystyrene (PS) ellipsoids with an aspect ratio (AR) of about 3 at the air/water interface, produced using the thermo-mechanical stretching technique[54] with different fabrication protocols. In the fabrication process, spherical PS colloidal particles (radius $R_0$ = 0.55 µm) are embedded in either a poly(vinyl alcohol) (PVA) or poly(vinyl pyrrolidone) (PVP) films (Fig. 1a).[54] This composite film is stretched in an oil bath above the glass transition temperature of the polymeric particles, deforming them into ellipsoids (Fig. 1b). The resulting anisotropic particles are subsequently retrieved by dissolving the films in water, followed by cleaning in either an isopropanol (IPA)/water mixture or pure water (Fig. 1c).

The synthesized ellipsoids are then spread at an air/water interface using ethanol as spreading agent, and their self-assembly behaviour is observed in situ using optical microscopy. A striking difference is observed depending on the fabrication protocol. Ellipsoidal particles prepared following the original work by Ho et al.,[54] using a PVA film and cleaning in IPA/water mixtures, self-assemble into a "chain-like" structure, where the ellipsoids are in direct contact in a side-to-side configuration (Fig. 1d). The same ellipsoids but cleaned using dissolution and subsequent centrifugation/redispersion in water, self-assemble into a non-close-packed structure where the ellipses are visibly separated from each other. They assume both side-to-side and tip-to-tip configurations (Fig. 1e). Ellipsoids prepared in a PVP film, on the other hand, assemble into a "flower-like" tip-to-tip configuration (Fig. 1f), but also visibly separated and non-close-packed.

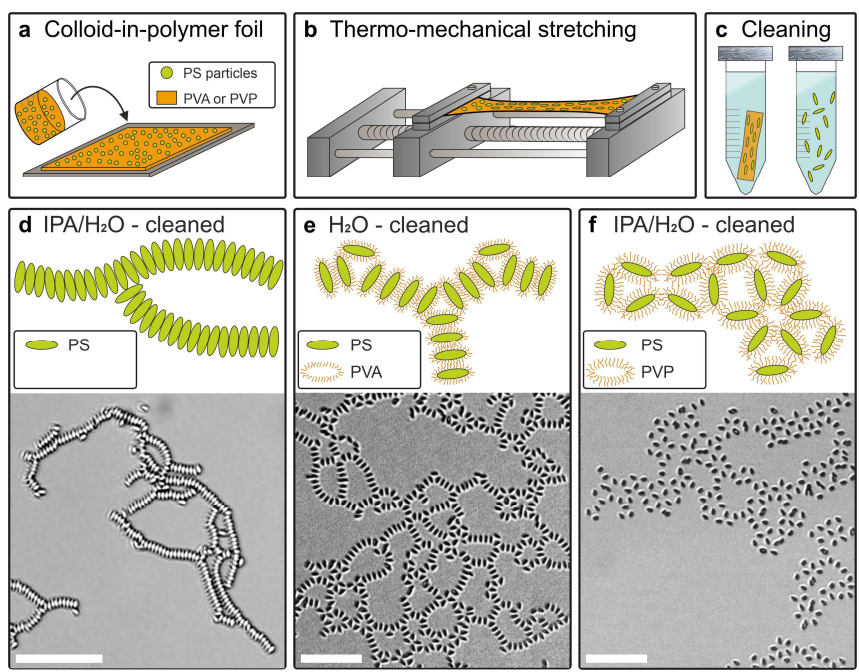

**Figure 1: The self-assembly behaviour of ellipsoids at the air/water interface depends on their fabrication process.** a-c) Schematic illustration of the fabrication process: a) Colloidal polystyrene (PS, radius $R_0$ = 0.55 µm) particles are embedded in a PVA or PVP film, followed by (b) thermo-mechanical stretching. c) The ellipsoidal particles are recovered by dissolving the film in $H_2O$ followed by cleaning using centrifugation and redispersion in either pure $H_2O$ or IPA/$H_2O$ mixtures. d-f) Schematic illustration and optical microscopy images of the interfacial self-assembly at an air/water interface for ellipsoidal particles prepared by different fabrication procedures. d) Ellipsoids prepared in a PVA film and cleaned in an IPA/$H_2O$ mixture assemble in direct contact in a "chain-like" side-to-side arrangement. e) The same ellipsoids but cleaned solely with $H_2O$ assemble in a mixture of side-to-side and tip-to-tip configurations and retain a non-close packed arrangement. f) Ellipsoids prepared in a PVP film assemble in a "flower-like" tip-to-tip configuration with non-close packed arrangement, even when cleaned with an IPA/$H_2O$ mixture. Scale bars: 20 µm.

We attribute the difference in self-assembly behaviour to the presence of polymeric chains adsorbed onto the particle surface during the fabrication process. An IPA/water mixture is required to sufficiently remove PVA from the ellipsoid surfaces.[54,55] In contrast, using pure water for the centrifugation/redispersion leaves a layer of PVA chains on the particle surface with a swollen thickness of roughly 40 nm (Fig. S1), as determined by dynamical light scattering.[55] These PVA chains extend at the liquid interface to reduce the surface energy and thereby form a 2D corona around the ellipsoidal particles, which is even visible in situ using cryo-scanning electron microscopy (SEM).[47] This prevents the particles from coming into close contact and, as we will discuss later, affects the energy landscape of the tip-to-tip and side-to-side configurations.

Assembling PVA-coated ellipsoids at liquid interfaces is challenging. Typically, particles are spread at the liquid interface using ethanol or isopropanol as spreading agents. However, when these spreading agents are mixed with the particle dispersion, PVA is readily removed from the particle surface, making it difficult to control their behaviour at liquid interfaces. To overcome this challenge, we replaced the PVA matrix used in the stretching process with PVP (polyvinylpyrrolidone). PVP polymer chains remain at least partially attached to the ellipsoidal particle surface even after cleaning with an IPA/water mixture (Fig. 1f, Fig. S1). These ellipsoids thus retain their polymer shell after spreading at the air/water interface, evidenced by the non-close-packed arrangement at the liquid interface (Fig. 1f) and thus allow a systematic and reliable study of their self-assembly as a function of aspect ratio and shell-to-core ratio. Therefore, from this point forward, core-shell ellipsoids with a PVP shell are used in this manuscript.

In bulk liquid, both PVA (166 kDa) and PVP (360 kDa) chains produce a similar shell thickness of around 40 nm (Fig. S1). However, ellipsoids with a PVP shell appear more separated at the air/water interface compared to those with a PVA shell (cf. Fig. 1e, Fig. 1f). We hypothesize that PVP chains extend further at the liquid interface, leading to a more extended corona around the ellipsoids than PVA chains. Although the exact mechanism is not yet fully understood, our data suggest that the physicochemical properties of the polymer shell significantly influence corona formation at liquid interfaces, which, in turn, affects the self-assembly behaviour of the ellipsoidal particles.

**The interfacial self-assembly behaviour of core-shell ellipsoids is influenced by their aspect ratio and shell-to-core ratio.** We next investigate how the self-assembly behaviour of these ellipsoids varies with aspect ratio (Fig. 2). Spherical polymethyl methacrylate (PMMA) particles (radius $R_0$ = 1.65 μm) remain fully separated (Fig. 2a), indicating predominant repulsive interactions. In contrast, when stretched into ellipsoidal particles, they exhibit attractive interactions at the liquid interface (Fig. 2b-h). At low aspect ratios, ellipsoids primarily form a "flower-like" tip-to-tip assembly (Fig. 2b-d). As the aspect ratio increases, "flower-like" and "chain-like" structures with side-to-side arrangement coexist (Fig. 2d-f). These ellipsoids form network-like structures where tip-to-tip triangles form branching points of individual side-by-side chains. The length of the chains continuously increases, until, at the highest aspect ratios (AR>6), the predominant assembly is in the form of chains.

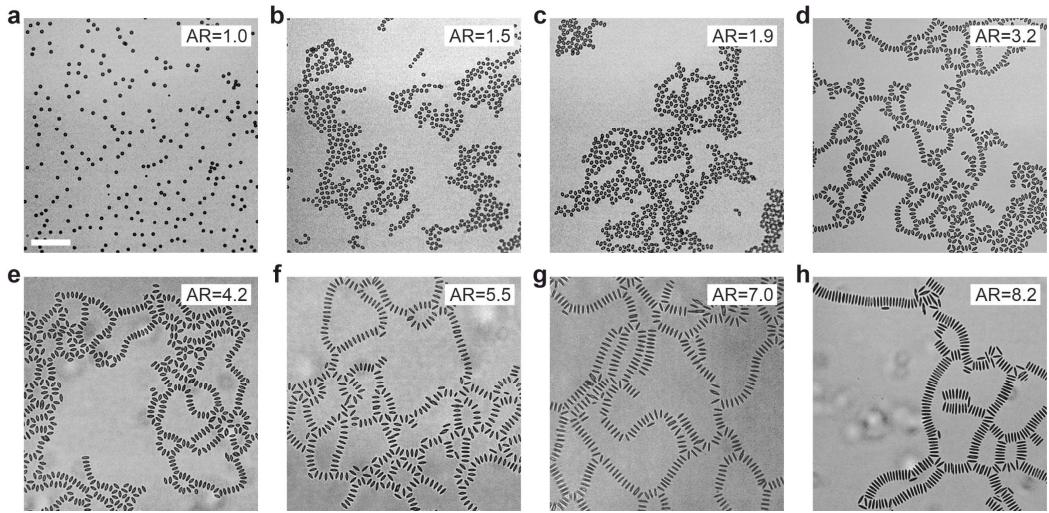

**Figure 2: Interfacial self-assembly behaviour of core-shell ellipsoids as a function of aspect ratio (AR).** Microscopy images of the self-assembly behaviour of (a) spherical particles and (b-h) ellipsoids with increasing aspect ratio (AR). We observe a transition from a "flower-like" tip-to-tip assembly (b-d) to a "chain-like" side-to-side assembly (e-h). Scale bar: 50 μm.

Next, we quantify the interfacial self-assembly behaviour of the ellipsoidal particles as a function of their shell-to-core ratio and aspect ratio (Fig. 3, Fig. 4). Experimentally, we vary the shell-to-core ratio, $h/R_0$, by stretching colloidal particles with different initial radii ($R_0 = 0.55$ μm and $R_0 = 1.65$ μm), while maintaining the same PVP shell polymer (Fig. 3, Fig. S2). The interfacial shell thickness $h$ formed by the PVP polymer chains stretched at the liquid interface is quantitatively determined by averaging over the distances between ellipsoids in the side-to-side configuration (Fig. S3). This model assumes that the ellipsoidal particles in this configuration are in shell contact, which seems reasonable as capillary forces will tend to bring the particles into the closest possible configuration. The measurements in Fig. S3 confirm that $h/R_0$ remains independent of the aspect ratio. It is important to note that the PVP shell thickness in bulk is roughly 40 nm, but once adsorbed to the liquid interface, the PVP polymer chains extend to form a micron-sized 2D shell, also termed a corona, along the interface as they spread out to reduce surface tension.[55,57]

To quantify the self-assembled structures at the interface, we employ a U-Net deep-learning model combined with the watershed algorithm to obtain instance-level segmentations of the particles.[58,59] These segmentations are used to extract the position, size, and orientation of the particles through ellipse fitting. We then classify the particle configurations as either tip-to-tip (green), side-to-side (orange) or unclassified (purple) by comparing the distance and orientation of each ellipse to its nearest neighbours. Ill-defined particles (black) are excluded from the analysis. The color-coded outlines of the ellipses are then overlayed onto the microscopy images (Fig. 3, top rows). A detailed description of the segmentation and categorization protocol is provided in the Methods section.

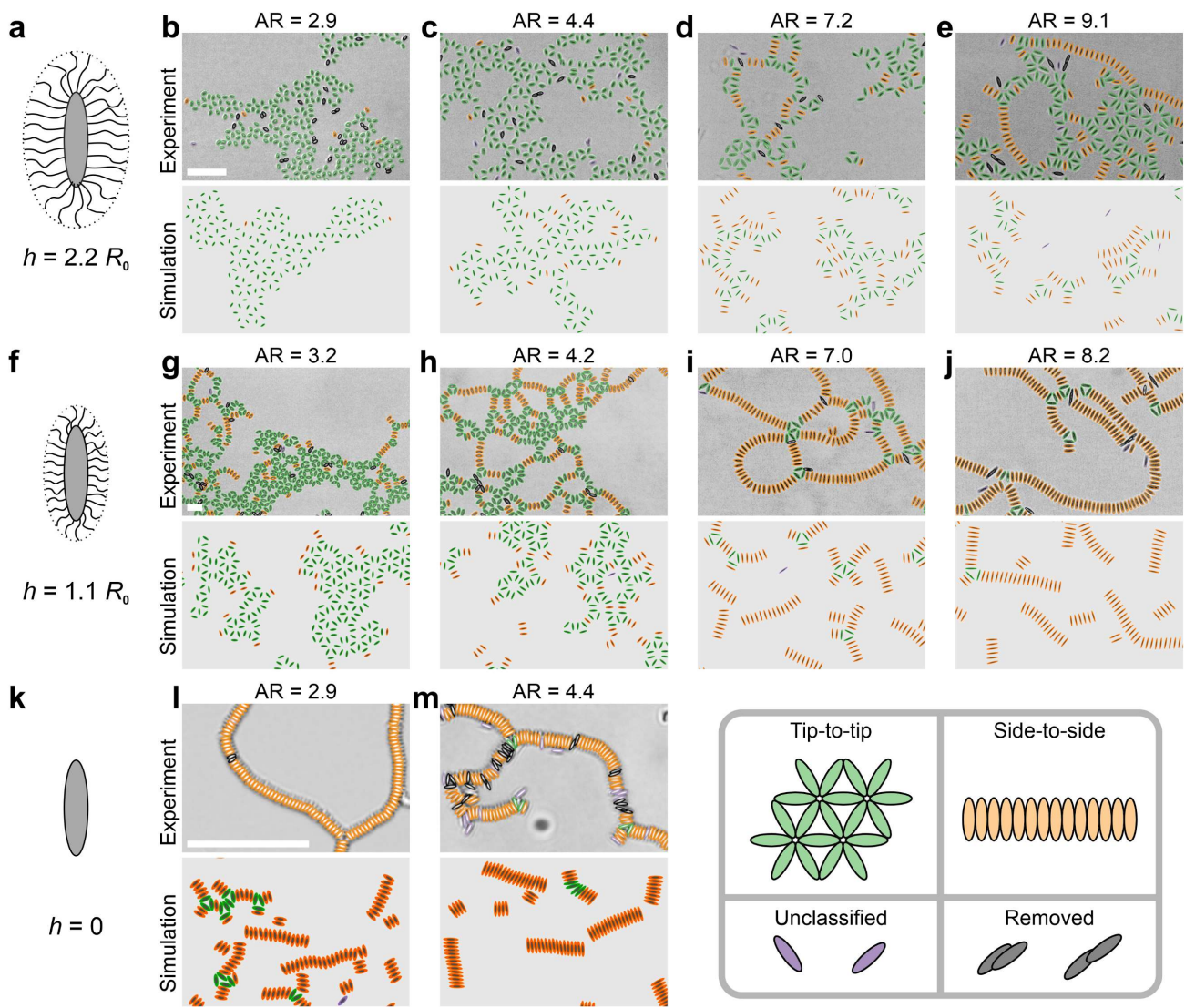

**Figure 3: Self-assembly of core-shell ellipsoids at the air/water interface as a function of shell thickness $h$ and aspect ratio AR.** a,f,k) Schematic illustration of the core-shell ellipsoids. b-e, g-j,l-m) Microscopy images (top) with colour-coded overlay and corresponding Monte Carlo simulations (bottom). Ellipsoids in tip-to-tip are colour-coded green, in side-to-side orange, unclassified purple and ill-defined, removed ellipsoids in black. a-e) Ellipsoids with a shell thickness $h = 2.2\ R_0$ predominantly assemble in tip-to-tip configuration at low and intermediate aspect ratios (b,c) and coexistence of tip-to-tip and side-to-side at higher aspect ratio (d,e). f-j) Ellipsoids with a shell thickness $h = 1.1\ R_0$ assemble in a tip-to-tip configuration at low aspect ratio (g), a mixture of tip-to-tip and side-to-side at intermediate aspect ratio (h) and side-to-side at high aspect ratio (i,j). k-m) Ellipsoids without a shell ($h = 0$) only assemble side-to-side. Scale bars: 20 µm.

We find that ellipsoids with the higher shell-to-core ratio (Fig. 3a-e, $h = 2.2\ R_0$), produced using polystyrene particles with an initial radius $R_0 = 0.55$ µm, predominantly assemble in a "flower-like" tip-to-tip structure at low and intermediate aspect ratios (Fig. 3b,c). At higher aspect ratios, there is a notable coexistence with a "chain-like" side-to-side structure (Fig. 3d,e). A similar pattern is observed for ellipsoids with a lower shell-to-core ratio (Fig. 3f-j, $h = 1.1\ R_0$), where tip-to-tip structures are present at low aspect ratios (Fig. 3g), coexistence is seen at intermediate aspect ratios (Fig. 3h), and side-to-side configurations dominate at higher aspect ratios (Fig. 3i, j). However, the transition to side-to-side configurations occurs at lower aspect ratios for these ellipsoids. For ellipsoids without a shell (Fig. 3k-m, $h = 0$), essentially only "chain-like" side-to-side assemblies are observed experimentally (Fig. 3l,m), irrespective of aspect ratio. These trends are quantitatively confirmed through a statistical analysis of the structures formed, as shown in Fig. 4.

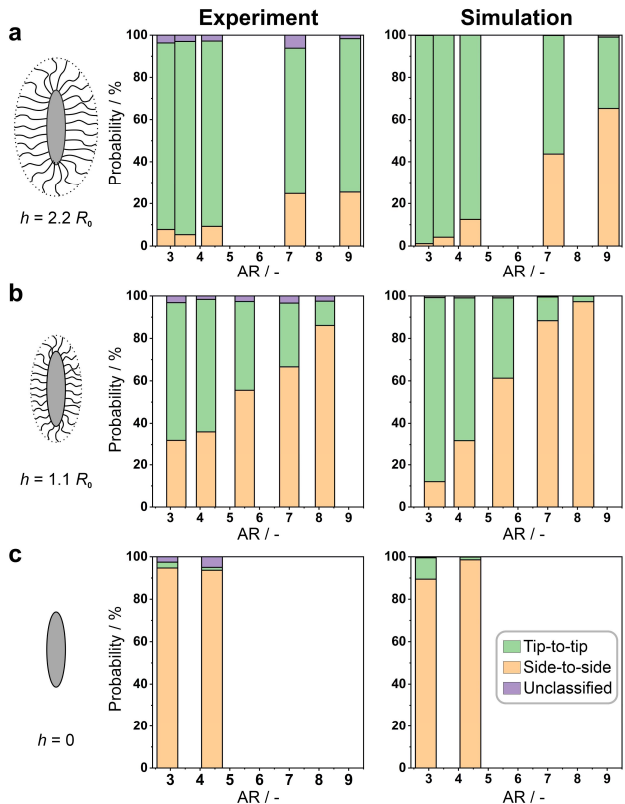

**Figure 4: Quantitative analysis of the self-assembly behaviour of core-shell ellipsoids.** Probability of finding ellipsoids in a tip-to-tip (green) or side-to-side (orange) configuration. For core-shell ellipsoids (a,b), there is a continuous transition from a tip-to-tip assembly to a side-to-side assembly with increasing aspect ratio. c) Ellipsoids without a shell almost exclusively assemble into side-to-side configurations.

**Monte Carlo simulations corroborate the experimental self-assembly behaviour.** To gain a deeper understanding of the fundamental forces driving the self-assembly of core-shell ellipsoids at the liquid interface, we perform Monte Carlo (MC) simulations of core-shell ellipsoids with aspect ratios and shell thicknesses $h$ corresponding to the experimental system, including both steric and capillary interactions between the particles. For simplicity, we neglect electrostatic repulsions.

As discussed earlier, the PVP chains adsorbed to the ellipsoid particle core are stretched out more significantly at the liquid interface compared to the bulk. The steric interactions between the core-shell ellipsoids are therefore primarily due to the 2D shell formed by the interfacial PVP chains rather than the 3D shell formed by the bulk PVP chains. In refs.[57,60], we studied the self-assembly of core-shell particles at liquid interfaces that were subjected to an external compression where it is sometimes energetically favourable for the shells of neighbouring particles to locally collapse so that their cores come into contact. In contrast, the core-shell particles we are studying here are not subjected to any external compression and the steric repulsion between the 2D shells is sufficient to prevent particles from coming closer than shell-shell contact, as experimentally evidenced by the formed non-close-packed patterns (Fig. 1, Fig. 2). In the experiments, we also observe that this shell repulsion occurs at both the sides and the tips of particles and, for simplicity, we assume that the shell thickness is the same at the sides and the tips. In our MC simulation, we therefore treat the core-shell ellipsoids as hard ellipses with long and short axis lengths of $a' = a + h$ and $b' = b + h$ respectively, where $h$ is the experimentally measured shell thickness which is independent of the aspect ratio AR (Fig. S3). From the conservation of the core volume, the long and short axis lengths of the core are given by $a = R_0 \text{AR}^{2/3}$, $b = R_0 \text{AR}^{-1/3}$ respectively. Specifically, we reject any MC move that causes particles to come closer than the contact separation between the hard ellipses $\sigma_c$, where $\sigma_c$ is calculated using the Berne-Pechukas model,[61] see Supplementary Information.

The capillary interactions between the core-shell ellipsoids come from the quadrupolar deformation of the three-phase contact line on the ellipsoid surface due to the constant contact angle requirement at the contact line.[37] Note that the contact line resides on the PVP-coated core rather than the boundary of the shell, as we assume that the shell is essentially a 2D object formed by interfacially adsorbed polymer chains. In addition, it is reasonable to assume that the PVP shell renders the core more hydrophilic and we therefore assign a water contact angle of the core as $\theta_w = 40°$ in our model.[47] To model the capillary interactions between the core-shell ellipsoids, we treat each ellipsoid core as a capillary quadrupole in elliptical coordinates[35,62] and the resultant expression for the capillary interaction potential is given in Supplementary Information. The advantage of working in elliptical coordinates is that it allows us to accurately model the capillary interactions between rod-like particles using a small number of capillary multipoles,[35] in our case one multipole at quadrupolar order. To verify the accuracy of the model, we compare the model (solid lines) to Surface Evolver simulations (data points) for the capillary interaction energy $V$ as a function of the interparticle separation $r_{12}$ in Fig. 5a and Fig. S5. We compare ellipsoids with no shells in the side-to-side configuration (orange) and tip-to-tip configuration (green) for lower aspect ratio ellipsoids where the numerical simulations are stable. Note that in these figures, we have used $H_e$, the amplitude of the elliptical quadrupole, as a fitting parameter to fit the model to the Surface Evolver data. The fitted values of $H_e$ for different aspect ratio AR are given in Table S1. The model captures the key features of the numerical data almost quantitatively, including the far-field quadrupolar scaling of $V \sim -1/r_{12}^4$ (dashed black line) and the near-field deviations from this scaling, confirming that modelling the contact line undulations as elliptical quadrupoles is a good approximation. In particular, the model correctly predicts that the lowest energy configuration for two ellipsoids with no shells is where the ellipsoids are in side-to-side contact rather than tip-to-tip contact (note that we plot $-V$ in the vertical axis of Fig. 5a and Fig. S5).

The MC simulations were performed at a core area fraction of $\eta = \pi ab/8a'^2$ to ensure that the system is in the dilute regime (i.e., core-shell ellipsoids on a hexagonal lattice can freely rotate about their centres without interfering with each other) like in the experiments. The energy scale for capillary interactions is $\gamma H_e^2$, where $\gamma$ is the interfacial tension of the air-water interface (see Supplementary Information). The importance of capillary interactions relative to the thermal energy is therefore characterized by the normalized temperature $T^* = k_B T/\gamma H_e^2$ and in Table S1 in Supplementary Information, we show that the experimental system is effectively in the low temperature regime $T^* \ll 1$. To model the experimental system, we therefore quench the system from a high initial temperature to a final temperature of $T^* = 0.2$. The choice of the final temperature represents a good compromise between being low enough for the system to be in the low temperature regime and still being high enough for the MC simulation to equilibrate the system efficiently. Further details of the MC and Surface Evolver simulations can be found in Methods.

In Fig. 3, we show final snapshots from the MC simulations of core-shell particles for different aspect ratios and shell thicknesses (bottom rows) and directly compare them to their experimental counterparts (top rows). The MC simulation accurately reproduces the experimentally observed trends, namely that increasing aspect ratio favours the formation of "chain-like" side-to-side arrangements, while increasing shell thickness favours "flower-like" tip-to-tip configurations. We quantitatively analyse the MC simulation snapshot using the same statistical analysis as for the experimental data (see Methods for further details). In Fig. 4, we quantitatively compare the statistical classification of the different structures formed in the MC simulations as a function of the aspect ratio and shell thicknesses with their experimental counterparts. The data confirms that the MC simulation reproduces the experimentally observed trend that increasing aspect ratio favours the formation of "chain-like" side-to-side arrangements, while increasing shell thickness favours "flower-like" tip-to-tip formations.

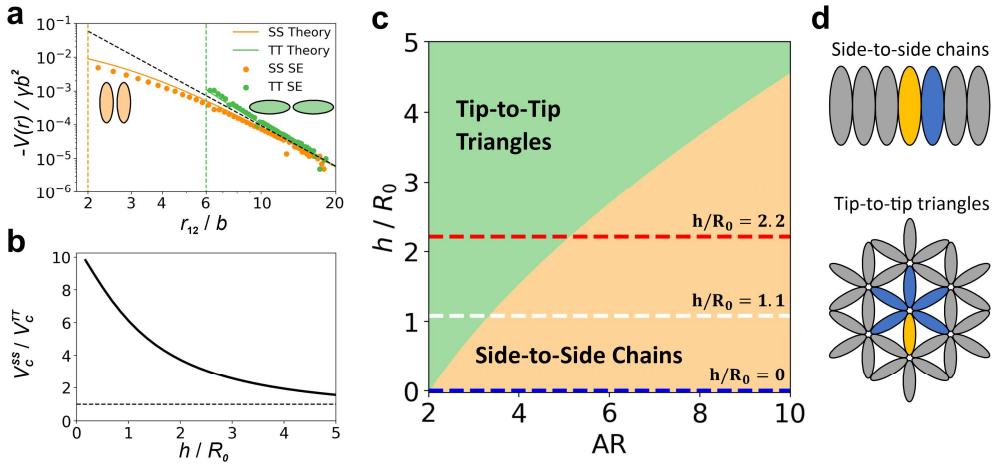

**Figure 5: Theoretical evaluation of interactions between core-shell ellipsoids at liquid interfaces.** a) Capillary interaction energy between two bare ellipsoids (i.e., no shell) with aspect ratio 3 as a function of centre-to-centre separation $r_{12}$ normalized by the length of the short axis of the ellipsoids, $b$, for ellipsoids oriented side-to-side (orange) and tip-to-tip (green). The lines are calculated analytically assuming the ellipsoids are elliptical quadrupoles while the points are calculated numerically using Surface Evolver. Note that we are plotting $-V$ on the vertical axis. The orange and green dashed vertical lines denote the centre-to-centre separation for side-to-side and tip-to-tip contact respectively while the black dashed line represents the quadrupolar power law. b) Ratio of capillary energies for core-shell ellipsoids in side-to-side vs. tip-to-tip contact as a function of shell thickness $h$ calculated from analytical theory assuming the cores are elliptical quadrupoles. Note that this ratio is always greater than unity (horizontal black dashed line) for all $h$ values studied. c) Zero temperature phase diagram for core-shell ellipsoids in the aspect ratio (AR) vs. shell to core ratio $h/R_0$ plane. The experimental values of $h/R_0$ are indicated by the blue, white and red dashed lines, respectively. d) Minimum energy configurations of side-to-side chains and tip-to-tip triangular lattice that are used to compute the phase diagram, where the energy per particle is calculated from the interaction between the yellow particle and its neighbouring blue particles.

We note from the snapshots in Fig. 3 that for each aspect ratio and shell thickness, the MC simulations not only reproduce the proportion of side-to-side chains versus tip-to-tip triangular lattices but also the local microstructure of the particle clusters and networks that are observed experimentally. One feature of the microstructure that is not captured so well by the simulations is that the cluster size of side-to-side chains tend to be smaller than what is observed experimentally, especially for the higher aspect ratio ellipsoids. We attribute this discrepancy to two reasons. First, the interaction potential for side-to-side contacts is lower than for tip-to-tip contacts and this energy difference increases with increasing aspect ratio so that the relaxation time for side-to-side chains is much longer compared to tip-to-tip triangular lattices in the low temperature regime for larger aspect ratio ellipsoids. Second, the area fraction we use in the simulations roughly scales as $\sim AR^{-1}$ and therefore decreases with increasing aspect ratio AR. These factors mean that for the same overall simulation length, the higher aspect ratio systems are not as well equilibrated as the lower aspect ratio systems, resulting in smaller side-to-side chain lengths and cluster sizes for higher aspect ratio systems. However, apart from this discrepancy, the agreement between simulation and experiments for both the statistics and the local microstructure of the self-assembled structures is remarkable, especially considering the minimalistic assumptions of our theoretical model. The fact that our simulations capture the key features of the experiments suggests that the self-assembly of core-shell ellipsoids is primarily driven by a competition between steric repulsions from the shell and capillary interactions from the core, while electrostatic interactions seem to play a negligible role.

**Minimum energy calculations rationalize the transition between tip-to-tip and side-to-side arrangement.** To assess the thermodynamics underpinning the observed self-assembled structures, we first consider how the interaction potential between core-shell ellipsoids varies with aspect ratio and shell thickness. As we increase the aspect ratio of the core-shell ellipsoids, side-to-side contacts become increasingly more energetically favourable compared to tip-to-tip contacts, essentially because the difference in centre-to-centre separation between the two configurations increases, see Fig. S5 in SI. This explains why increasing aspect ratio favours the formation of side-to-side chains compared to tip-to-tip triangular lattices.

In Fig. 5a, we focus on the role of shell thickness on the assembly process, using ellipsoidal particles with an aspect ratio of 3 as a model. We first note that as we increase the shell thickness $h$, the fractional increase in the centre-to-centre separation $r_{12}$ will be greater for ellipsoids in side-to-side contact compared to those in tip-to-tip contact. Since from Fig. 5a, the capillary interaction between the ellipsoidal cores is to a first approximation an inverse power law in $r_{12}$ (black dashed line), this means that the capillary bond energy (i.e., the magnitude of the attractive potential at contact) for core-shell ellipsoids in side-to-side contact should drop more than those in tip-to-tip contact as we increase shell thickness. This point is confirmed in Fig. 5b where we plot the ratio $V_c^{SS}/V_c^{TT}$ as a function of shell thickness $h$ for core-shell ellipsoids with aspect ratio AR = 3, where $V_c^{SS}$ and $V_c^{TT}$ are the capillary bond energies for side-to-side and tip-to-tip contacts respectively calculated from the elliptical quadrupole model. We see that the capillary bond energy for side-to-side contacts falls relative to that for tip-to-tip contacts as we increase $h$. Despite this decrease, we note that $V_c^{SS}/V_c^{TT} > 1$ for the shell thicknesses explored in this study, i.e., the capillary bond energy for side-to-side contact is always greater than that for tip-to-tip contact, so that this effect alone is not enough to tip the balance in favour of the formation of tip-to-tip structures.

However, in both the experiments and the simulations, the core-shell ellipsoids with large $h$ do not form tip-to-tip chains but "flower-like" tip-to-tip triangular lattices, where the number of capillary bonds per particle is greater than in the chain state. Specifically, the number of capillary bonds per particle is 1 for side-to-side chains and 5 for tip-to-tip triangular lattices, as schematically shown in Fig. 5d by the interactions between a central particle (yellow) and its neighbours (blue). In Fig 5c, we quantify the total capillary interactions for side-to-side and tip-to-tip arrangements as a function of aspect ratio and shell thickness, taking into account both the evolution of $V_c^{SS}/V_c^{TT}$ with shell thickness, and the increased number of neighbours for triangular tip-to-tip lattices. These calculations are based on the elliptical quadrupole model[35,62] (see Supplementary Information), where the centre-to-centre separation between nearest neighbours in shell-to-shell contact is calculated using the analytic formula given in ref. [63].

In our calculations, we assume the structures are infinite, i.e., we neglect defects and edge effects. We also neglect interactions of the yellow particle with particles beyond the blue particles, i.e., we assume an effective cutoff distance of $\approx a' = a + h$ for particle interactions; this is a good approximation since we are considering quadrupolar interactions which fall off rapidly with separation. Finally, we assume that the equilibrium structure for any given geometry is the structure with the minimum energy per particle. This assumption is justified since the experimental system is in the low temperature regime $T^* \ll 1$ where entropic effects are negligible. Fig. 5c shows the resultant zero-temperature phase diagram for core-shell ellipsoids in the aspect ratio AR-shell thickness $h$ plane.

This phase diagram accurately reproduces the trends observed in both experiments and simulations. Increasing the aspect ratio favours the formation of side-to-side chains, while increasing shell thickness favours the formation of tip-to-tip triangular lattices. In particular, it predicts that as we increase shell thickness from $h/R_0 = 0$ via $h/R_0 = 1.1$ to $h/R_0 = 2.2$ (blue, white and red horizontal lines, respectively), the transition from tip-to-tip triangular lattices to side-to-side chains occurs at higher and higher aspect ratios. Note that we do not see pure phases but a coexistence of different phases in both the experiments and the simulations, indicating that the system is not globally ergodic. However, the fact that the equilibrium structures predicted by the zero-temperature phase diagram correlate well with the dominant structures seen in both experiments and simulations suggests that the structures shown in Fig. 5d are kinetically accessible and the experimental core-shell ellipsoid system is at least locally ergodic.

## Conclusions

We investigate the self-assembly of polymeric ellipsoidal particles fabricated through thermo-mechanical stretching at a liquid interface, with a particular focus on the role of soft shells formed from polymeric residues during fabrication. When these polymers on the ellipsoid surface adsorb at a liquid interface, they spread under the influence of surface tension, forming a 2D shell around the ellipsoid that prevents direct contact between neighbouring ellipsoids. We demonstrate that this alteration in interfacial structure significantly changes the minimum energy landscape and, consequently, the self-assembly behaviour of these ellipsoidal particles.

Using a combined experimental and simulation approach, we systematically explore the resultant self-assembly behaviour and identify two key trends: increasing the aspect ratio of the ellipsoids favours their assembly into a "chain-like" side-to-side configuration, while increasing the shell thickness promotes "flower-like" tip-to-tip configurations. A phase diagram, based on the total capillary interactions as a function of core separation and the number of neighbours, reveals a transition in minimum energy structures from side-to-side to tip-to-tip configurations with increasing shell thickness and decreasing aspect ratio, corroborating both experimental and simulation findings.

Our results may explain contradictory reports on the self-assembly of elliptical particles by accounting for the significant effects of the often overlooked polymer corona on interactions at a liquid interface. The proposed framework is general and may provide a foundation for the predictive self-assembly of anisotropic particles into predetermined, complex arrangements. This can be achieved through the rational engineering of polymer coronae as two-dimensional spacers separating particles at a liquid interface.

## Methods

### Materials

All chemicals were obtained from commercial sources and used as received if not otherwise stated. Acrylic acid (99 %, Sigma Aldrich), ammonium persulfate (APS, 98%, Sigma Aldrich), azobisisobutyronitrile (>97 %, VWR), PVP K90 (360'000 kD, Sigma Aldrich), methyl methacrylate (>99%, Sigma Aldrich), methanol (99.8%, Sigma Aldrich), polyvinyl alcohol (PVA, Mw 146.000-186.000, 87-89 % hydrolyzed, Sigma Aldrich), ethanol (99.9%, Sigma Aldrich) and glycerol (99.5 %, Sigma Aldrich) were used as received.

Double-filtered and deionized water (18.2 MΩ·cm, double reverse osmosis by Purelab Flex 2, ELGA Veolia) was used throughout this study. Styrene as monomer (ReagentPlus with 4-tert-butyl catechol as a stabilizer, ≥99%, Sigma Aldrich) was washed with 10 wt-% NaOH aqueous solution to remove the inhibitor. The monomer was subsequently purified by flash column chromatography with activated $Al_2O_3$ (basic 90 for column chromatography, Carl Roth) using nitrogen gas.

**PS particle synthesis:** PS particles with a radius of 0.55 μm were synthesized using a surfactant-free emulsion polymerization method. In a 500 mL triple-neck round-bottom flask, 250 mL of water was heated to 80°C and degassed by bubbling nitrogen gas for 30 minutes. Then, 80 g of styrene and 0.4 g of the comonomer acrylic acid, dissolved in 5 mL of water, were added under constant stirring. After 5 minutes, 0.1 g of APS, dissolved in 5 mL of water, was introduced. The reaction was carried out at 80°C for one day. Following this, the mixture was cooled to room temperature, filtered, and purified through a series of centrifugation, redispersion, and dialysis against water for two months.

For the synthesis of smaller PS particles with a radius of 0.19 μm (used to measure the polymer shell thickness via dynamic light scattering (Malvern Zetasizer Nano ZS, Fig. S1), a similar protocol was followed with adjusted reactant concentrations. Specifically, 10 g of styrene, 0.1 g of acrylic acid, and 0.1 g of APS were used.

**PMMA particle synthesis:** PMMA particles with a radius of 1.65 µm were synthesized using a dispersion polymerisation method. In a one litre bottle, 45 g of PVP K90 (360'000 kD) is dissolved in 400 g of methanol. The fluorescent comonomer 7-nitrobenzo-2-oxa-1,3-diazole-methyl methacrylate was prepared according to the instructions of Jardine and Bartlett.[64] 0.375 g of this comonomer was dissolved in 15 g of methanol. Once these ingredients were fully dissolved, they were transferred to a one litre three-necked round-bottomed flask. Into this flask was also added 27.5 g of methyl methacrylate and 46 g of distilled water. The necks of the flask were fitted with a stirrer, a nitrogen inlet and a condenser. The flask was placed in an oil bath and the reaction mixture brought to reflux (72 °C). Whilst adjusting this flask to reflux, 0.375 g of the initiator azobisisobutyronitrile was dissolved in 10 g of methyl methacrylate. Once reflux had been achieved, the initiator system was added and the reaction allowed to proceed for 24 hours. After this time the reaction mixture was allowed to cool to room temperature and the particles were cleaned by centrifugation and redispersion in excess methanol six times. For the seventh step, the centrifugation was repeated and then the particles were dispersed in the minimum amount of methanol possible and then excess water was added to the dispersion. A further three centrifugation and redispersion steps were completed except this time distilled water was used as the suspending fluid. The particles were kept in distilled water prior to use.

**Ellipsoid fabrication:** Ellipsoidal particles were prepared using the established method by Ho et al.[54] For pure ellipsoids (without any adsorbed polymers) and PVA-coated ellipsoids, 6.75 g of PVA is dissolved in 100 mL of distilled water heated to 90 °C under continuous stirring. After the PVA dissolves, the solution is cooled to room temperature, and 5 mL of a 2 wt% PS particle dispersion is added. For PVP-coated particles, 10 g of PVP is dissolved in 100 mL of distilled water at 90 °C under stirring. After cooling, 1.1 g of glycerol is added to improve the ductility of the matrix. Then, 2 mL of a 13 wt% PMMA dispersion or 8 mL of a 2 wt% PS particle dispersion is added. Both the PVA and PVP particle mixtures are spread into thin films on lacquered aluminium plates in a fume hood and dried for 24 hours.

The dried films were cut into 2 x 8 cm$^2$ strips and fixed into a custom-made stretching device. The PVP films were additionally heated in an oven at 60 °C to ensure complete evaporation of any entrapped water. A household fryer equipped with a stirrer was used as an oil bath. For PVA matrices, the oil bath was set to 135 °C, while for PVP matrices, it was set to 160 °C. The fixed film was placed in the oil bath for five seconds, followed by variable stretching depending on the desired ellipsoid aspect ratios. The film was then cooled to room temperature and cleaned with a tissue soaked in isopropanol (IPA) to remove any remaining oil.

The middle part of the stretched film (approximately 5 cm) was cut out and dissolved in water. If not stated otherwise, the PVA and PVP-coated ellipsoids were all cleaned four times by centrifugation and redispersion in water. Pure ellipsoids were obtained by cleaning the ellipsoids stretched in PVA matrices an additional three times with a 3:7 IPA/water mixture, followed by three times with water using centrifugation and redispersion. The same procedure was used for PVP-coated ellipsoids in Fig. 1f. The arithmetic mean long and short axes of the ellipsoids were determined by measuring 100 particles in SEM images (GeminiSEM 500, Carl Zeiss AG, Germany).

**Interfacial self-assembly and optical microscopy imaging:** The experiments were performed using video microscopy (CCD camera, IDS UI-3060CP-M-GL R2) using a custom-built inverted optical microscope. We obtained different magnifications by using either a 20x or a 40x Nikon air immersion objective. Two glass slides (Marienfeld, size 1.5), in one of them was a 5 mm circular hole cut, were glued together to create a sample cell. The cell was filled with 5 µL of water. The aqueous ellipsoid dispersion was mixed with 15 vol% ethanol as spreading agent and 0.1 µL of the dispersion was spread at the liquid interface. For each sample, at least 10 images were taken.

**Surface Evolver:** The capillary interaction energy and the amplitude of contact line undulations were calculated using Surface Evolver, a finite element method which represents the interface as a mesh of triangles and displaces vertices to minimize energy subject to constraints.[65] An adjustable triangular mesh between $0.02b - 0.1b$ with quadratic edge lengths was used to capture the shape of the interface near the contact line more accurately. Since the interactions between the ellipsoids are essentially quadrupolar and fall off rapidly with separation, a relatively small simulation box with reflecting boundary conditions at $x = \pm 6a$ and $y = \pm 6a$ was used. To calculate the capillary interaction potential between two ellipsoids, the reflecting walls to calculate the interfacial energy of a single ellipsoid interacting with its image at a reflecting wall was exploited, varying the particle surface-to-reflecting wall distance from $0.1b$ to $3a$ and using the ellipsoid at the centre of the simulation box to represent the 'infinite' separation case.

**Monte Carlo simulations:** NVT Metropolis simulations on an ensemble of 625 core-shell ellipsoids with periodic boundary conditions were performed using a fixed rectangular box with aspect ratio 2:√3 with particles initially arranged in a hexagonal array with random azimuthal orientations and a core area fraction $\eta = \pi ab/8(a+h)^2$ to ensure that the system is in the dilute regime. Each MC move consisted of a simultaneous translation and rotation move, with an adjustable maximum translational distance of $d_{max}$ and azimuthal rotation angle of $d_{max}/a$ about the particle centre to ensure an acceptance probability of 30% for the MC moves. The particles were initially randomized at a high temperature of $T^* = 100$ for $10^3$ attempted moves per particle, then quenched to $T^* = 0.2$ for a further $10^6$ attempted moves per particle. To reduce computation time, particle interactions for separations greater than a cutoff distance of $2a + 10b$ were neglected.

**Particle detection and measurements:** The positions and orientations of the particles were determined using a deep learning approach.[58,59] First, a U-Net predicted the distance transform map of the segmentation of each particle, normalized per particle such that the maximum value of the distance transform map is 1. The distance transform map was preferred as a target over immediately predicting the segmentation map since it better distinguished closely packed particles.

Once the distance transform map is acquired, instance segmentations of the particles were determined. Two binary segmentations were calculated by comparing the distance transform map to two simple thresholding criteria, the values of which were determined using a hyperparameter sweep on a holdout set after training. The first, sparser binarization was used to detect the individual particles. The second binarization was used as a panoptic foreground segmentation of the image. Finally, the watershed algorithm was applied to the two binarizations to acquire instance segmentation maps.

Next, orientation and the minor and major axes of the particles were estimated using the Python library *scipy* applied to the individual instance segmentations. The minor and major axes were additionally used as a sanity check, lightly filtering detections that are too far outside the range of expected sizes.

**Neural network data annotation and training:** To efficiently collect sufficient annotated data for training the artificial neural network (ANN), an active learning approach was used. First, each experimental image was split into 128x128 pixel² patches with a 50 % overlap. Initially, a uniformly sampled selection of 20 such patches was presented to an oracle to annotate using an ad-hoc user interface by drawing ellipses over the particles in view.

From this, an initial training dataset was constructed using the Python libraries *Deeplay* and *DeepTrack*.[58,59] First, since the samples are monodisperse, we normalized the sizes of all annotated ellipses to the median ellipse for each experiment. The target image was computed as the normalized distance transformation map of each ellipse. In the case of overlapping ellipses, the maximum value of the overlapping ellipses' distance transform map for each conflicting pixel was used. During training, the images were augmented using integer multiples of 90° rotations, random horizontal mirroring, and a high pass filter with a random cutoff frequency.

Using this initial dataset, a committee of five ANNs was trained for 20 epochs each, using $L_1$ loss, a batch size of 32, and an Adam optimizer with a learning rate of 0.001. This committee was subsequently used to query the most valuable patches for further annotations by the oracle. First, the prediction of each ANN on each image was computed. Second, the patch was sampled using the following scheme: i) a particle type was randomly chosen to ensure a diverse sampling; ii) for each patch not yet annotated, the disagreement within the committee as the average pixel-wise variance was calculated; iii) one patch at random from the top 10 % scoring patches was selected to alleviate overfitting the criterion.

The scheme was repeated until a desired budget was exhausted. The budget was set to 100 patches. A validation set was created in parallel, where each patch was sampled uniformly from all not yet annotated patches for a given experiment. This ensured that the validation results were not biased by the training data sampling criterion.

For each selected patch, the oracle was presented by the ellipses acquired by the process described by the method outlined in the previous section. The user tuned this prediction by deleting false positive ellipses and adding missed ellipses. This increased annotation significantly since most patches had only a few errors. The annotated data was added to the training pool, and a new committee was trained. This process was repeated until desired F1-score was reached on the validation-set. Once sufficient data had been annotated for the desired quality, a final ANN was trained using the same training parameters as the committee, but for a total of 100 epochs.

**Statistical analysis of clusters:** Particles were classified as belonging to side-to-side chains or tip-to-tip triangular clusters based on their separation and orientation relative to other particles in the system, with the criteria based on these quantities chosen to ensure that the classification coincides with the classification based on visual inspection. For this analysis, the centre-to-centre separation between two nearest-neighbour core-shell ellipsoids in contact was $d_C = 2b'$ and $d_T = \sqrt{a'^2 + 3b'^2}$ respectively for the side-to-side chains and tip-to-tip triangular lattices shown in Fig. 5d,[63] where $a' = a + h$, $b' = b + h$. For the simulation system, a particle was classified as belonging to a side-to-side chain if its centre-to-centre separation to any other particle satisfied the condition $r_{12} < 1.3 d_C$ and the relative orientation of the two particles satisfied the condition $|\hat{a}_1 \cdot \hat{a}_2| \geq 0.9$, where $\hat{a}_1, \hat{a}_1$ are the unit vectors along the semi-major axes of the two particles. On the other hand, a particle as belonging was classified to a tip-to-tip triangular cluster if its centre-to-centre separation to any other particle satisfied $0.7 d_T \leq r_{12} \leq 1.15 d_T$ and the relative orientation of the two particles satisfied $|\hat{a}_1 \cdot \hat{a}_2| < 0.9$. If a particle belonged to both a chain and a triangular cluster, we classified it as belonging to a triangular cluster.

For the experimental system, a neural network was used to analyse the experimental micrographs to capture the position of the ellipsoid centres, the long and short axes lengths and the orientation of the long axis relative to one of the lab frame axes. The data was then filtered by eliminating any particles whose long axis length is smaller than 60% of the mean or two standard deviations greater than the mean. Further, any overlapping particles were removed by assuming that all ellipsoids have a long and short axes length equal to the mean and all particle pairs whose centre-to-centre separation is smaller than the contact separation predicted by the Berne-Pechukas model[61] were also removed from the analysis (see Supplementary Information). We note that the neural network underpredicts the aspect ratio of the ellipsoidal particles compared to values obtained from direct measurements of the experimental micrographs but fortunately this discrepancy does not have a significant impact on our statistical classification. Having filtered the experimental data as described above to remove any artefacts of the ellipsoid identification process, an ellipsoid was classified as being part of a side-to-side chain if its separation and orientation relative to any other particle satisfied the conditions $r_{12} < 1.3 d_C$, $|\hat{a}_1 \cdot \hat{a}_2| \geq 0.9$ and $|\hat{a}_1 \cdot \hat{r}_{12}| < 0.5$, where $\hat{r}_{12}$ unit vector along the centre-to-centre line connecting the two particles. On the other hand, an ellipsoid was classified as being part of a tip-to-tip triangular cluster if its separation and orientation relative to any other particle satisfied the conditions $0.5 d_T \leq r_{12} \leq 0.8 d_T$, $|\hat{a}_1 \cdot \hat{a}_2| < 0.9$ and $|\hat{a}_1 \cdot \hat{r}_{12}| > 0.2$. If an ellipsoid belonged to both a chain and a triangular cluster, we classified it as belonging to a triangular cluster. The statistical classification above was carried out for $4-20$ experimental micrographs for each aspect ratio and shell thickness.

**Corresponding Author:**

mrey@uni-muenster.de

**Author Contributions:**

*These authors contributed equally.

**Conflict of interests:**

There are no conflicts to declare.

**Funding Sources:**

M.R. acknowledges funding from Sklodowska-Curie Individual Fellowship (Grant No.101064381). We further acknowledge the use of the Cryo FIB/SEM bought with the EPSRC grant EP/P030564/1. JLE, DMAB and NV acknowledge funding from the European Union's Horizon 2020 research and innovation programme under grant agreement No 861950, project POSEIDON. JLE and DMAB acknowledge the Viper High Performance Computing facility of the University of Hull. NV acknowledges funding of the German Research Council (Deutsche Forschungsgemeinschaft) under grant number VO 1824/6-2.

Supplementary Figures:

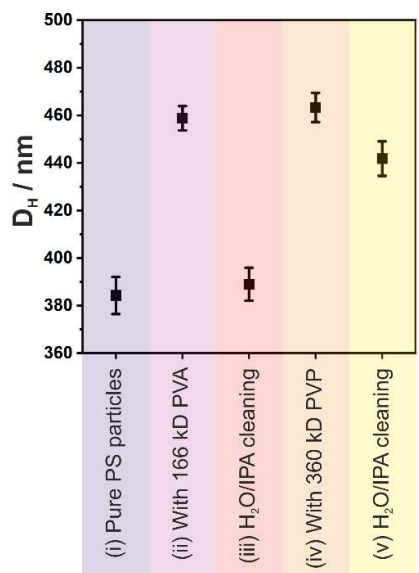

**Figure S1: Hydrodynamic diameter ($D_H$) for the respective particle dispersions measured with dynamic light scattering.** (i) Pure spherical polystyrene (PS) particles. (ii) After mixing PS particles with 166 kD PVA and subsequent removal of non-adsorbed PVA by centrifugation and redispersion in water, an increase in the $D_H$ is observed, indicating the formation of a polymeric shell of physisorbed PVA. (iii) Washing these particles with an IPA/$H_2O$ mixture removes the physisorbed PVA, as evidenced by the decrease in $D_H$, returning to the value of the pure PS dispersion. (iv) When PS particles are mixed with 360 kDa PVP and the excess PVP is removed by centrifugation and redispersion in water, the hydrodynamic diameter increases, similar to the PVA case, indicating a physisorbed polymer shell. (v) After cleaning with the IPA/$H_2O$ mixture, the $D_H$ remains higher than that of the pure PS particles, suggesting that the polymer shell at least partially persists even after washing.

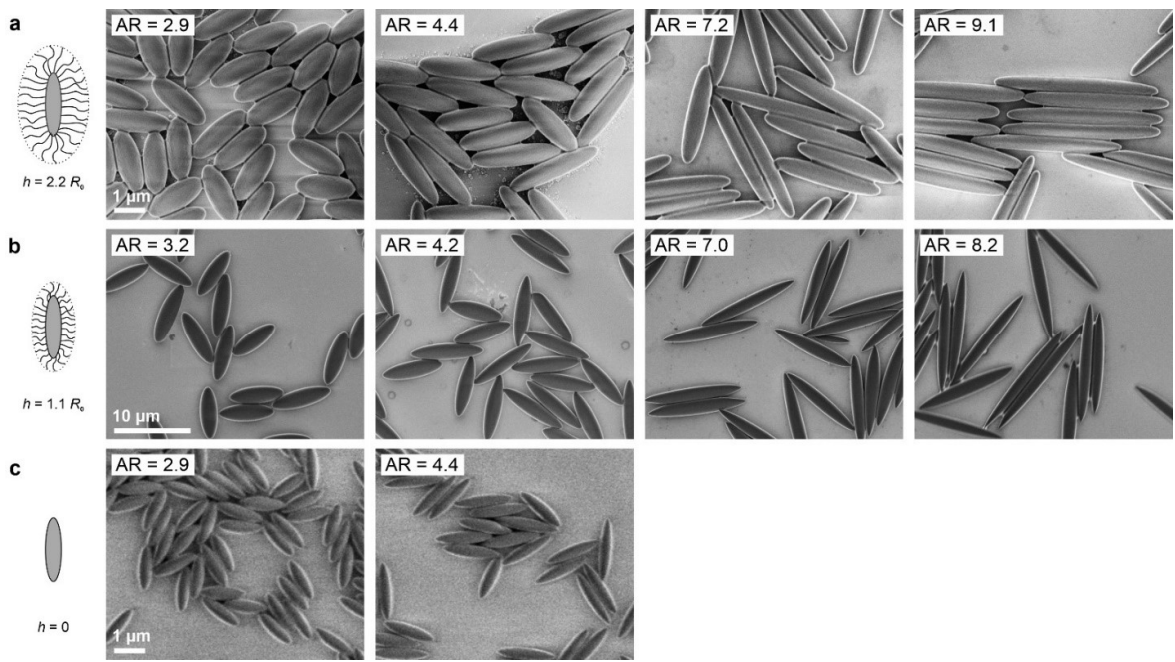

**Figure S2: Scanning electron microscopy (SEM) images of the ellipses used in Fig. 3.** a) PS ellipsoids containing a PVP shell. b) PMMA ellipsoids with a PVP shell. c) Pure PS ellipsoid with no shell.

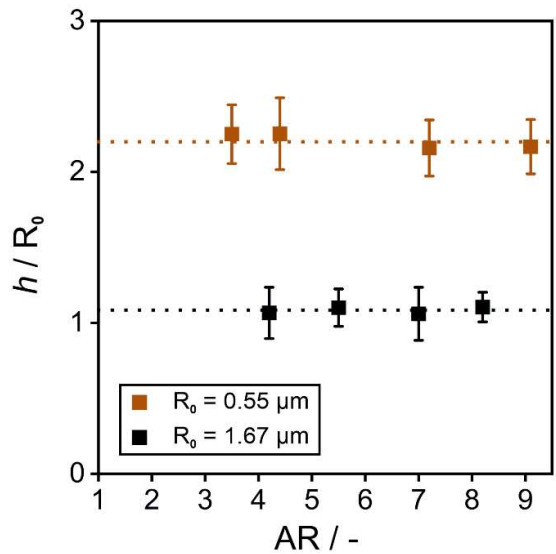

**Figure S3: Shell thickness *h* as a function of aspect ratio.** The shell thickness *h* is determined by measuring the center-to-center distance for ellipsoids in a side-to-side arrangement from optical microscopy images and subtracting the corresponding short-axis measurements obtained from SEM images.

**Steric Interaction between Core-Shell Ellipsoids**

As discussed in the main paper, to model steric interactions between core-shell particles in our Monte Carlo (MC) simulations, we assume the shells are impenetrable and treat the core-shell particles at liquid interfaces as hard ellipses with long and short axis lengths of $a' = a + h$ and $b' = b + h$ respectively, where $h$ is the experimentally measured shell thickness, $a = R_0 \text{AR}^{2/3}$, $b = R_0 \text{AR}^{-1/3}$ are the long and short axis lengths of the ellipsoidal core respectively, $R_0$ is the radius of the original unstretched spherical core and $\text{AR} = a/b$ is the aspect ratio of the final stretched ellipsoidal core. This approach for modelling steric repulsions requires us to know the contact separation $\sigma_c$ between hard ellipses, which is given in the Berne-Pechukas model [1] by

$$\sigma_c(\hat{u}_1, \hat{u}_2, \hat{r}) = \sigma_\perp \left( 1 - \frac{1}{2} \left[ \frac{(\hat{r} \cdot \hat{u}_1 + \hat{r} \cdot \hat{u}_2)^2}{1 + \chi \widehat{\hat{u}_1 \cdot \hat{u}_2}} + \frac{(\hat{r} \cdot \hat{u}_1 - \hat{r} \cdot \hat{u}_2)^2}{1 - \chi \widehat{\hat{u}_1 \cdot \hat{u}_2}} \right] \right)^{-1/2}. \quad (S1)$$

where $\hat{u}_1, \hat{u}_2$ are the unit vectors along the long axis of the two interacting ellipsoids, $\hat{r}$ is the unit vector along the line joining the particle centres,

$$\chi = \frac{\sigma_\parallel^2 - \sigma_\perp^2}{\sigma_\parallel^2 + \sigma_\perp^2}$$

and $\sigma_\parallel = 2(a + h)$, $\sigma_\perp = 2(b + h)$ are the length and width of the hard ellipse respectively. Note that for prolate ellipsoids, the Berne-Perchukas model for $\sigma_c$ is accurate to within 2%. [2]

**Capillary Interaction between Core-Shell Ellipsoids**

As discussed in the main paper, we assume that the capillary interactions between the core-shell ellipsoids are due to quadrupolar contact line undulations on the ellipsoidal cores and this interaction is a function of the centre-to-centre separation of the interacting ellipsoids $r_{12}$ and the azimuthal angle that the long axis of each particle makes to the centre-to-centre $\varphi_1, \varphi_2$ as shown in Figure S4. In this section, we follow ref. [3] and calculate the capillary interaction energy between two ellipsoid cores $V(r_{12}, \varphi_1, \varphi_2)$ by treating the contact line undulations on the cores as capillary quadrupoles in elliptical coordinates.

Assuming particle-centred coordinate systems and that the $x$ and $y$ axis of the Cartesian coordinate system are aligned along the long and short axis of the ellipsoid respectively in the interfacial plane, the elliptical coordinates $(s, t)$ are related to the Cartesian coordinates $(x, y)$ by the transformations

$$x = \alpha \cosh(s) \cos(t) \quad (S2)$$

$$y = \alpha \sinh(s) \sin(t) \quad (S3)$$

where $s, t$ are analogous to radial and polar coordinates respectively in circular polar coordinates and $\alpha$ is a scale factor (with dimensions of length) that determines the eccentricity of the constant-$s$ lines in the elliptical coordinate system. Following ref. [3], we assume that the projection of the contact line on the interfacial plane corresponds to the outline the ellipsoidal core projected on the interfacial plane is given by the ellipse $s = s_0$. Making these assumptions, one can show that

$$\alpha = b\sqrt{\text{AR}^2 - 1} \quad (S4)$$

$$s_0 = \tanh^{-1}(1/\text{AR}). \quad (S5)$$

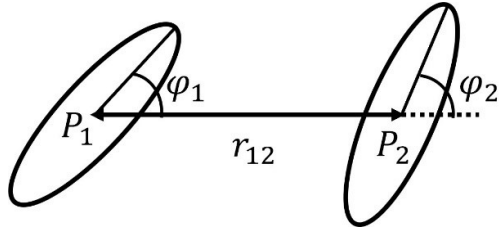

**Figure S4:** The coordinates specifying the configuration of two interacting ellipsoids at a liquid interface. The ellipses represent the projection of the contact line on each ellipsoid on the interfacial plane while $P_1$, $P_2$ denote the centres of the two ellipsoids on the interfacial plane.

A key parameter in calculating the capillary interaction is the position of the centre of particle 2 (P2) in a coordinate system where the centre of particle 1 (P1) is the origin. If we denote the position of P2 in Cartesian coordinates as $(x', y')$, from Figure S4, we see that $x' = r_{12} \cos \varphi_1$, $y' = -r_{12} \sin \varphi_1$. Substituting this result into Equations (S2), (S3) and rearranging, we can write the position of P2 in elliptical coordinates $(s^*, t^*)$ in terms of $r_{12}$ and $\varphi_1$ as

$$s^* = \sinh^{-1}\left\{\left[\frac{1}{2}\left(\frac{r_{12}^2}{\alpha^2} - 1\right) + \frac{1}{2}\left[\left(\frac{r_{12}^2}{\alpha^2} - 1\right)^2 + \frac{4 r_{12}^2 \sin^2 \varphi_1}{\alpha^2}\right]^{1/2}\right\}^{1/2}\right] \quad (S6)$$

$$t^* = \sin^{-1}\left[\frac{-r_{12} \sin \varphi_1}{\alpha}\left\{\frac{1}{2}\left(\frac{r_{12}^2}{\alpha^2} - 1\right) + \frac{1}{2}\left[\left(\frac{r_{12}^2}{\alpha^2} - 1\right)^2 + \frac{4 r_{12}^2 \sin^2 \varphi_1}{\alpha^2}\right]^{1/2}\right\}^{-1/2}\right]. \quad (S7)$$

Finally, the interaction energy between the two elliptical quadrupoles is given by ref. [3]

$$V(r_{12}, \varphi_1, \varphi_2) = -2\gamma\pi\alpha^2 H_e \cosh(s_0) \sinh(s_0) [C'_{11} \cos(2\varphi_1 - 2\varphi_2) - C'_{21} \sin(2\varphi_1 - 2\varphi_2)] \quad (S8)$$

where $\gamma$ is the interfacial tension of the fluid interface, $H_e$ is the amplitude of the elliptical quadrupolar and

$$C'_{11} = \frac{H_e e^{2s_0} e^{-2s^*}}{r_{12}^2 \cos^2 \varphi_1 (\tanh^2 s^* - \tan^2 t^*)^2}\left[4\cos(2t^*)(\tanh^2 s^* - \tan^2 t^*) - 8\sin(2t^*)\tan t^* \tanh s^*\right.$$
$$+ 2(\tanh^2 s^* + \tan^2 t^*)(\cos(2t^*)\tanh s^* - \sin(2t^*)\tan t^*) - \frac{2\cos 2t^* \tanh s^*}{\cosh^2 s^*}$$
$$\left. - \frac{2\sin 2t^* \tan t^*}{\cos^2 t^*} + \frac{4(\cos(2t^*)\tanh s^* - \sin(2t^*)\tan t^*)}{\tanh^2 s^* + \tan^2 t^*}\left(\frac{\tanh^2 s^*}{\cosh^2 s^*} - \frac{\tan^2 t^*}{\cos^2 t^*}\right)\right] \quad (S9)$$

$$C'_{21} = \frac{H_e e^{2s_0} e^{-2s^*}}{r_{12}^2 \cos^2 \varphi_1 (\tan^2 t^* + \tanh^2 s^*)^2}\left[4\sin(2t^*)(\tanh^2 s^* - \tan^2 t^*) + 8\cos(2t^*)\tan t^* \tanh s^*\right.$$
$$+ 2(\tanh^2 s^* + \tan^2 t^*)(\sin(2t^*)\tanh s^* + \cos(2t^*)\tan t^*) - \frac{2\sin 2t^* \tanh s^*}{\cosh^2 s^*}$$
$$\left. + \frac{2\cos 2t^* \tan t^*}{\cos^2 t^*} + \frac{4(\sin(2t^*)\tanh s^* + \cos(2t^*)\tan t^*)}{\tanh^2 s^* + \tan^2 t^*}\left(\frac{\tanh^2 s^*}{\cosh^2 s^*} - \frac{\tan^2 t^*}{\cos^2 t^*}\right)\right] \quad (S10)$$

where $s^*$, $t^*$ are given by Equations (S6), (S7).

Note that in Equations (S8) – (S10), we have corrected some typographical errors in the corresponding equations in ref. [3]. Specifically, in Equation (S8), there is a minus between the two terms inside the square bracket (not a plus as shown in ref. [3]), while in Equations (S9), (S10), the denominator on the right hand side contains an extra factor of $\cos^2 \varphi_1$ (which is missing in ref. [3]). Note also that, in spite of this extra factor, $C'_{11}$, $C'_{21}$ do not diverge for $\varphi_1 \to \pm \pi/2$ but tend towards finite limiting values. To avoid any numerical instabilities in our calculations, we therefore set $C'_{11}$, $C'_{21}$ to be equal to these limiting values for $\varphi_1$ very close to $\pm \pi/2$.

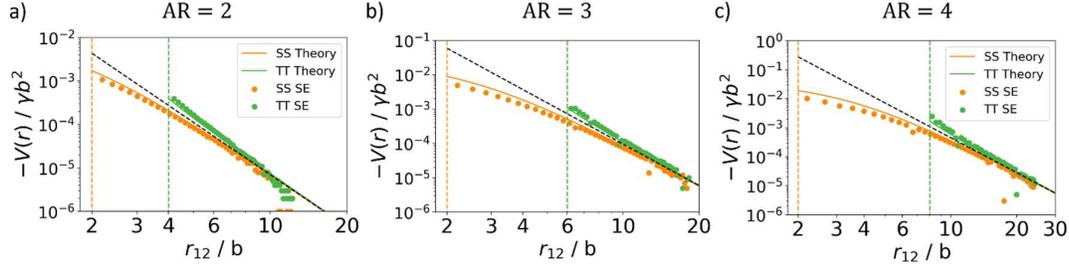

**Figure S5:** Comparison of elliptical quadrupole model with Surface Evolver simulations for the capillary interaction energy as a function of centre-to-centre separation $r_{12}$ for ellipsoids with no shells and contact angle $\theta_w = 40°$ with aspect ratio (a) AR = 2; (b) AR = 3; (c) AR = 4. Note that in all cases we plot $-V$ in the vertical axis. We compare theory (solid lines) to simulations (data points) for ellipsoids in the side-to-side configuration (red) and tip-to-tip configuration (blue). The red and blue vertical dashed lines represent the contact separation for the side-to-side and tip-to-tip configuration respectively while the dashed black line represents the quadrupolar power law. The values of the fitting parameter $H_e$ used to fit theory to simulation for the different AR are given in Table S1.

In Figure S5a,b,c, we compare our theoretical model, i.e., Equations (S6)-(S10), with Surface Evolver simulations for the interaction energy as a function of $r_{12}$ for ellipsoids with no shells and contact angle $\theta_w = 40°$ which have aspect ratios AR = 2,3,4 respectively. Specifically, we compare theory (solid lines) to simulations (data points) for ellipsoids in the side-to-side (orange) and tip-to-tip configuration (green). We use $H_e$ as a fitting parameter to fit the theory to the simulation data and the fitted values of $H_e$ for the different AR are given in Table S1. We see that the theoretical model captures the key features of the numerical data almost quantitatively, including the far-field quadrupolar scaling of $V \sim -1/r_{12}^4$ (dashed black lines) and the near-field deviations from this scaling. The fitted values of $H_e$ also agree with the amplitude of contact line undulations $H_0$ calculated directly from Surface Evolver simulations of isolated ellipsoids to within a factor of around 2, where we define the amplitude to be $H_0 = (z_{max} - z_{min})/2$, i.e., half the difference in the maximum height $z_{max}$ and minimum height $z_{min}$ of the contact line (see Table S1). The good agreement between theory and simulations in Figure S5 and Table S1 confirms that modelling the contact line undulations on the ellipsoid cores as elliptical quadrupoles is a good approximation.

As discussed in the main paper, the importance of capillary interactions relative to the thermal energy is characterized by the normalized temperature $T^* = k_B T / \gamma H_e^2$. In Table S1, we list the values of $T^*$ for ellipsoids with AR = 2,3,4 where we assume that $R_0 = 0.55 \ \mu m$, $T = 300$ K and $\gamma = 70$ mN·m$^{-1}$. We see that in all cases $T^* \ll 1$ and that $T^*$ decreases with increasing AR. Since all the experimental ellipsoids have AR > 2.5 and $T^*$ will be lower for ellipsoids with larger $R_0$ (since $H_e$ is proportional to $R_0$), we conclude that all experimental systems we are studying in this paper are in the low temperature regime where $T^* \ll 1$.

**Table S1:** The amplitude of quadrupolar contact line undulations for ellipsoids with $\theta_w = 40°$ and different aspect ratios AR obtained from fitting the elliptical quadrupole model to the Surface Evolver simulation data

in Figure S5 ($H_e$) and calculated directly from Surface Evolver simulations of isolated ellipsoids ($H_0 = (z_{max} - z_{min})/2$, where $z_{max}$, $z_{min}$ are the maximum and minimum heights respectively of the contact line). We also list the normalized temperature $T^* = k_B T/\gamma H_e^2$ for the ellipsoids assuming $R_0 = 0.55$ μm, T = 300 K and $\gamma = 70$ mN·m$^{-1}$.

| Aspect Ratio AR | $H_e$ | $H_0$ | $T^*$ |
|---|---|---|---|
| 2 | $0.0205b$ | $0.0506b$ | $7.4 \times 10^{-4}$ |
| 3 | $0.0445b$ | $0.0878b$ | $2.1 \times 10^{-4}$ |
| 4 | $0.0690b$ | $0.115b$ | $1.0 \times 10^{-4}$ |